\documentclass[aps,reprint,twocolumns,pra,showpacs,superscriptaddress,floatfix,notitlepage,nofootinbib]{revtex4-1}
\usepackage{amssymb,amsmath,amsfonts,eucal} 
\usepackage{amsthm}
\usepackage{mathtools}
\usepackage{physics}
\usepackage{array}
\usepackage{graphicx,epsfig}
\usepackage[colorlinks=true]{hyperref}

\begin{document}

\title{Distribution of the Ratio of Consecutive Level Spacings for Different Symmetries and Degrees of Chaos}

\author{\'{A}ngel L. Corps}
   \email[]{angelo04@ucm.es}% Your name
    \affiliation{Departamento de Estructura de la Materia, F\'{i}sica T\'{e}rmica y Electr\'{o}nica, Universidad Complutense de Madrid, Av. Complutense s/n, E-28040 Madrid, Spain}
\author{Armando Rela\~{n}o}
\email[]{armando.relano@fis.ucm.es}
\affiliation{Departamento de Estructura de la Materia, F\'{i}sica T\'{e}rmica y Electr\'{o}nica and GISC, Universidad Complutense de Madrid, Av. Complutense s/n, E-28040 Madrid, Spain}

\date{\today} % Leave empty to omit a date

\begin{abstract}
Theoretical expressions for the distribution of the ratio of consecutive level spacings for quantum systems with transiting dynamics remain unknown. We propose a family of one-parameter distributions $P(r)\equiv P(r;\beta)$, where $\beta\in[0,+\infty)$ is a generalized Dyson index, that describes the eigenlevel statistics of a quantum system characterized by different symmetries and degrees of chaos. We show that this crossover strongly depends on the specific properties of each model, and thus the reduction of such a family to a universal formula, albeit desirable, is not possible. We use the information entropy as a criterion to suggest particular ansatzs for different transitions, %Stringent numerical calculations with the classical random ensembles and real physical systems permits us to suggest a particular ansatz for all three Poisson-GOE, Poisson-GUE, and GOE-GUE transitions, 
with a negligible associated error in the limits corresponding to standard random ensembles. %This enables a convenient study of quantum chaos in a vast range of practical scenarios.
\end{abstract}

\maketitle

\section{Introduction}
Quantum chaos \cite{stockmann} has been of the utmost importance during a large period of time. The unraveling of how this behavior in the quantum realm emerges from classical mechanics would help us delve deeper into the correspondence principle. It would also cast light on new exotic phenomena. 

Classical chaos has been explored to a great extent and is solidly substantiated both  phenomenologically and mathematically \cite{strogatz}. The onset of classical chaos from the regular regime is unambiguously resolved by the KAM theory \cite{arnold}. %For the quantum counterpart, the uncertainty principle invalidates any characterization in terms of physical trajectories and there is an absence of functionally independent constants of motion \cite{neumann}. 
The pioneering work of Berry and Tabor \cite{berrytabor} states that for quantum Hamiltonians whose classical analog is integrable, the level statistics and their fluctuations properties follow a simple Poisson law. The transition from integrability to chaos is mediated by a universally shared dramatic change in the eigenlevel statistics \cite{gutzwiller}. 

The work of Bohigas, Giannoni, and Schmit \cite{bgs,bgscounter} meant the final link between random matrix theory (RMT) \cite{rmt} and the spectral fluctuation properties of quantum systems with a chaotic classical analog \cite{wigner1,wigner2,dyson}. Level fluctuations of quantum systems whose classical analogs are chaotic will fall into the descriptions of one of the three classical symmetry random ensembles: the Gaussian orthogonal, unitary and symplectic ensembles (GOE, GUE, GSE). Explanations in the semiclassical limit using the spectral form factor have been suggested \cite{demobgs,Saito2009,qgraphs1,qgraphs2,Gnutzmann2004,Sieber2001}. 
The conjecture has found an important amount of applications in several settings \cite{borgonovi,Gomez11,ullmo,rigol,Santhanam2001,Seba2003,Barthelemy2002,Kollath2010,Iyer2013,Collura2012, Janarek2018,Luitz2016}.

Random matrix ensembles describe energy levels of real systems at a statistical level within a local energy window in which the mean level density is set to unity. %For these statistical properties to be accurately compared with the predictions of RMT, the above requirement is of vital importance. 
For this, a transformation called unfolding needs to be performed. This consists in mapping the system eigenlevels by means of the smooth part of the density of states.
Therefore, knowledge of the system density of energy states is required. In principle, this quantity can be wildly dependent on each physical system. Additionally, an unfolded spectrum can suffer from numerous non-trivial spurious effects \cite{spurious,us}.  It is, then, desirable to seek alternatives for which the unfolding procedure plays no role. 

One of such tools, on which we focus in this work, is the \textit{distribution of the ratio of consecutive level spacings}, $P(r)$, which has been growing in popularity since its introduction to the scientific community \cite{defratios}, immensely propelled by the derivation of theoretical expected values for the Poisson, GOE, GUE, and GSE cases  \cite{ratios,ratios2} as well as the transparency of the analysis it provides. Probably its most interesting open issue is due to the fact that most physical systems cannot be fully taken into account by any of the standard regularity limits due to their intrinsic nature of \textit{intermediate} dynamics. This means that one needs to obtain results that apply when the \textit{degree of chaoticity} is not clear and needs to be assessed. So far, one model for the GOE-GUE transition has been exactly derived \cite{goegue}. For the Poisson-GOE transition, a heuristic suggestion for a particular system has also been proposed before \cite{kota1}, and an attempt to analytically solve the problem has been made as well \cite{kota2}.  Variants of this spectral statistic have been proposed and analyzed as well \cite{Kota2018,Bhosale2018,Tekur2018b,piotr2019}. In Ref. \cite{Tekur2018a}, exact and numerical results are provided to take into account neighboring localized states occurring in a typical quantum chaotic spectrum. The degree of chaos then depends on the coupling strength. 

Crucially, as we show in this work, \textit{ there cannot exist} a universal result that allows for the interpolation between regularity and symmetry classes for an infinite range of arbitrary systems. This seriously hinders the derivation of universally applicable formulae. Here, a practical ansatz is proposed relying on the information entropy as a criterion. 

This paper is organized as follows. In Sec. \ref{secII} we summarize the main results of our work. In Sec. \ref{secIII} we give details on the mathematical structure of our formula and show that it agrees with several known limits. In Sec. \ref{secIV} we show that our distribution reduces the error with respect to the theoretical GOE and GUE expressions given in \cite{ratios}, which we here extend. We also present scaling analysis of its parameters and error. In Sec. \ref{SecV} we  analyze the crossover from integrability to chaos in four different models and find that it is strongly system-dependent; we also analyze other kind of intermediate statistics. In Sec. \ref{secVI} we propose an ansatz derived from the information entropy. Finally, in Sec. \ref{SecVII} we gather the main conclusions.

\section{Summary of results}\label{secII}
For reference, we summarize here the practical results of our work. A detailed discussion can be found in subsequent sections. 

\textit{Crossover distribution}. We postulate the one-parameter distribution for the ratio of consecutive level spacings
\begin{equation}\label{distribution}
    %P(r)\equiv
   P_{\gamma\beta}(r)\equiv P(r;\beta,\gamma(\beta))=C_{\beta}\frac{(r+r^{2})^{\beta}}{\left[(1+r)^{2}-\gamma(\beta)r\right]^{1+3\beta/2}}.
\end{equation}
Here, $\beta\in[0,+\infty)$ is a Dyson-like index. The function $\gamma(\beta)$ is system dependent and determines the precise shape of the distribution. {\em It is not possible to find a universal $\gamma(\beta)$ covering all  crossovers from integrability to chaos for a general system}. However, it is possible to construct a practical ansatz. Finally, the normalization constant $C_{\beta}$ is calculated via the condition $\int_{0}^{\infty}\textrm{d}r\,P(r;\beta)=1$.

\textit{Constants and anstazs.} In Table \ref{info}, we summarize the main results for integrable and fully chaotic systems for our distribution Eq. \eqref{distribution}. It is worth remarking that our results for $\langle r\rangle $ and $\langle \widetilde{r}\rangle$ for GOE and GUE, where $\widetilde{r}$ is the  random variable with values $\widetilde{r}_{n}\equiv \min\{r_{n},1/r_{n}\}\in[0,1]$ with distribution $P(\widetilde{r})=2P(r)\Theta(1-r)$, are slightly different from the analytical results reached in \cite{ratios}. The latter are obtained from $3\times 3$ random matrices; ours introduce small corrections to better describe fits of numerical data. 

\begin{table}[h!]
\begin{center}
\setlength\extrarowheight{5pt}
 \begin{tabular}{|| c || c c c ||} 
 \hline
 Quantity & \hspace{0.4cm} Poisson & \hspace{0.4cm} GOE & \hspace{0.4cm} GUE \\ [0.5ex] 
 \hline\hline
$\beta$  & \hspace{0.4cm} $0$  & \hspace{0.4cm} $1$ & \hspace{0.4cm} $2$ \\[1ex]
 \hline
 $\gamma(\beta)$  & \hspace{0.4cm} $0$ & \hspace{0.4cm} ${\frac{4}{5}}$ & \hspace{0.4cm} ${\frac{8}{9}}$ \\[1ex]
 \hline
 $C_{\beta}$ & \hspace{0.4cm} $1$ & \hspace{0.4cm} ${\frac{96}{25}}$ & \hspace{0.4cm} $\approx12.6532$ \\ [1ex] 
 \hline
  $\langle {r}\rangle$ & \hspace{0.4cm} $\infty$ & \hspace{0.4cm} ${\frac{9}{5}}$ & \hspace{0.4cm} $\approx1.37584$\\ [1ex] \hline
 $\langle \widetilde{r}\rangle$ & \hspace{0.4cm} $2\ln2-1$ & \hspace{0.4cm} ${5-2\sqrt{5}
 }$ & \hspace{0.4cm} $\approx0.59769$\\ [1ex] \hline
\end{tabular}
\end{center}
\caption{Calculated values of the useful quantities $\beta$, $\gamma(\beta)$, $C_{\beta}$, $\langle r\rangle$, and $\langle \widetilde{r}\rangle$ for the crossover distribution Eq. \eqref{distribution} for Poisson, GOE, and GUE of dimensions $N\gg1$. }
\label{info}
\end{table}

In Table \ref{ansatzs}, we suggest practical ansatzs for Poisson-GOE, Poisson-GUE, and GOE-GUE crossovers. %As we will show later, they typically incur in insignificant errors so long as one considers around $10^{3}$ ratios. 
\begin{table}[h!]
\begin{center}
\setlength\extrarowheight{5pt}
 \begin{tabular}{||c c c||} 
 \hline
 Transition \hspace{0.0cm} & $\beta$ \hspace{0.0cm} & $\gamma(\beta)$ \\ [0.5ex] 
 \hline\hline
Poisson-GOE \hspace{0.0cm} & $0\leq\beta\lesssim 1$ \hspace{0.0cm} & $\displaystyle{0.80-1.69(1-\beta)+0.89(1-\beta)^{5}}$ \\[1ex]
 \hline
 Poisson-GUE \hspace{0.0cm} & $0\leq\beta\lesssim2$ \hspace{0.0cm} & $\displaystyle{0.92-1.42(2-\beta)+0.01(2-\beta)^{7}}$ \\[1ex]
 \hline
 GOE-GUE \hspace{0.0cm} & $1\lesssim\beta\lesssim2$ \hspace{0.0cm} & $\displaystyle{0.88-0.36(2-\beta)+0.28(2-\beta)^{3}}$ \\ [0.8ex] 
 \hline
\end{tabular}
\end{center}
\caption{Choice for $\gamma(\beta)$ for Poisson, GOE, and GUE crossovers from the information entropy.}
\label{ansatzs}
\end{table}

\section{Crossover distribution}\label{secIII}
We define the probability density function used in this work, detail the assumptions made in order to reach it, and briefly comment on some mathematical aspects.

The ratio of consecutive level spacings is a random variable $r$ taking on values \begin{equation}\label{ratio}
r_{n}\equiv \frac{E_{n+1}-E_{n}}{E_{n}-E_{n-1}},\,\,\,\,\forall n\in\{2,\dots,N-1\},\end{equation}
where $\{E_{n}\}_{n=1}^{N}$ is a complete set of energies in ascending order, that is, verifying $E_{n}\geq E_{m}$ whenever $n\geq m$. Since the distribution of $r$ and that of $1/r$ are the same \cite{ratios}, it follows that any probability density associated to this random variable must verify \begin{equation}\label{symmetry}
    P(r)=\frac{1}{r^{2}}P\left(\frac{1}{r}\right).
\end{equation} In a spirit similar to that of the Wigner surmise, a formula for the ratio distribution of two consecutive spacings was obtained in \cite{ratios} by analytically solving the $3\times3$ problem associated to the Poisson, GOE, GUE, and GSE cases. This probability density exhibits the same level repulsion as the nearest neighbor spacing distribution (NNSD) for vanishingly small values of $r$; explicitly, $P(r)\sim r^{\beta}$ for $r\to0$. We follow the same intuition to now suggest an expression that interpolates between different standard regularity classes and symmetries. We then ask our interpolating function to yield the correct theoretical limits when $\beta$ is fixed to the corresponding value. We propose, in analogy with the Brody distribution \cite{brody} for the NNSD, the probability density function
%\begin{equation}P:\,\,[0,+\infty)\to[0,+\infty)\end{equation} given by 
%\begin{equation}\label{distribution}
 %  P_{\gamma\beta}(r)\equiv P_{\gamma\beta}\left(r;\beta,\gamma(\beta)\right)=C_{\beta}\frac{(r+r^{2})^{\beta}}{\left[(1+r)^{2}-\gamma(\beta)r\right]^{1+3\beta/2}}.
%\end{equation}
given by Eq. \eqref{distribution}. Here, $\beta\in[0,+\infty)$ is taken to be a non-negative, continuous parameter. It can be thought of as a generalized Dyson-like index. The values $\beta=0,1,2,4$ correspond to Poisson, GOE, GUE, and GSE, respectively. The one-variable function $\gamma\equiv\gamma(\beta)$ uniquely establishes the maximum of $P(r)$ at each value of $\beta$. The analytical results of \cite{ratios} are recovered \textit{if} \begin{equation}\label{limits}
    \gamma(\beta=0)=0,\,\,\,\gamma(\beta=1,2,4)=1.
\end{equation} Here we note, however, that Eq. \eqref{limits} will \textit{not} be strictly fulfilled, since the original results were calculated from $3\times 3$ random matrices, and therefore deviations from Eq. \eqref{limits} are expected for larger systems. Results summarized in Table \ref{info} are then slightly different, but best suited for the typical matrix sizes of data analysis.
%Eq. \eqref{limits} guarantees the recovery of the known theoretical formulae for Poisson, GOE, GUE, and GSE \cite{ratios}--namely, Eq. \eqref{distribution} then simplifies to the Poissonian result 
%\begin{equation}\label{teoricas1}
%P_{P}(r;\beta=0)\equiv\frac{1}{(1+r)^{2}},
%\end{equation}
%and the Wigner-like results
%\begin{equation}\label{teoricas2}
%P_{W}(r;\beta=1,2,4)\equiv C_{\beta}\frac{(r+r^{2})^{\beta}}{(1+r+r^{2})^{1+3\beta/2}}.
%\end{equation}
%This ensures the compatibility of Eq. \eqref{distribution} with the exact probability distributions for the classical ensembles. Here we note, however, that Eq. \eqref{limits} will not be strictly fulfilled in practice, because these limits are expected to hold only for $3\times 3$ random matrices within the context of RMT. This is also the case for which Eqs. \eqref{teoricas1} and \eqref{teoricas2} are expected to hold true \cite{ratios}.

Since $P(r)$ is a probability density, it must verify $P(r;\beta,\gamma(\beta))\geq0$, $\forall r,\beta\in[0,+\infty)$. This leads to the condition 
\begin{equation}
  \gamma(\beta)<\min_{r\in[0,+\infty)}\frac{(1+r)^{2}}{r}=4,\,\,\forall\beta\in[0,+\infty),
\end{equation}
which in turn ensures the non-singularity of $P(r)$, $\forall r\in[0,+\infty).$ Here we draw attention to the nature of the $\gamma(\beta)$ function just defined. The decision to choose it as a one-variable function could seem arbitrary. However, since the transitions we will consider in this work are mediated by a single perturbative parameter, this is the choice that makes the most sense both physically and mathematically. %Otherwise, our model Eq. \eqref{distribution} would also depend on more than one parameter, while the regularity class of the systems we analyze will clearly change as the value of only one parameter is varied.% Any extra parameter dependence in our formula would then leave a conceptual gap with an information that corresponds to nothing in particular. 

Finally, the normalization constant $C_{\beta}$ is implicitly determined by the condition $\int_{0}^{\infty}\textrm{d}r\,P(r;\beta)=1$, $\forall \beta\in[0,+\infty)$. %Eq. \eqref{distribution} is the initial transiting model of our work. 

It is interesting to observe how Eq. \eqref{distribution} behaves asymptotically, which determines the structure of level repulsion \cite{explanation3}. In the domain $r\ll 1$, expanding at $r=0$ affords the Maclaurin representation
\begin{equation}\begin{split}\label{r0}
P(r) &\simeq C_{\beta}r^{\beta}\left[1+\left(-2-2\beta+\gamma+\frac{3\beta\gamma}{2}\right)r+\order{r^{2}}\right]\\ & =C_{\beta}r^{\beta}+\order{r^{\beta+1}},
\end{split}\end{equation}
as expected. Similarly, for $r\to\infty$ one has $P(r)\sim C_{\beta}r^{-(2+\beta)}$, which describes the distribution queue. 
%\begin{equation}\label{rinf}
%P(r)\sim C_{\beta}\frac{r^{2\beta}}{{r^{2+3\beta}}}=C_{\beta}r^{-(2+\beta)}.
%\end{equation}

The statistical moments given by Eq. \eqref{distribution} strongly depend on the value of $\beta$ and do not always exist. In particular, the $k$-th moment of the random variable $r$ is determined as 
\begin{equation}\begin{split}\label{moments}
\langle r^{k}\rangle_{\beta} & =\int_{0}^{\infty}\textrm{d}r\,r^{k}P(r;\beta)\sim \int_{0}^{\infty}\textrm{d}r\,r^{-(2+\beta-k)}<\infty\\ & \iff \beta>k-1,\,\,\forall\beta\in[0,+\infty),\,\,\,\forall k\in\mathbb{N}.
\end{split}
\end{equation}
%This means that our model does not yield a finite mean value $\langle r\rangle_{\beta=0}$ for Poisson, GOE does have a $\langle r\rangle_{\beta=1}$ but not variance, and so on.  
Thus, Eq. \eqref{distribution} successfully reproduces the same qualitative behavior with respect to the existence of moments as the original distribution for the classical random ensembles. In fact, Eq. \eqref{distribution} slightly improves on the mean values of $r$, when these exist, and those of $\widetilde{r}$ (see Table \ref{info} for details) with respect to the original ones \cite{ratios}.% Note, however, that as $\beta\to0$, $\langle r\rangle$ is badly conditioned, and for it to converge for values $0<\beta\ll 1$ very large statistics are called for. This inconvenience is of course avoided by $\langle \widetilde{r}\rangle$, which exhibits no singular behavior at any point.

Because no explicit expression for $\gamma(\beta)$ can be deduced, in what follows we perform non-linear fits that treat $\beta$ and $\gamma$ as unknown, independent parameters. Although $C_{\beta}$ can be numerically obtained via the normalization condition for $\gamma$ and $\beta$ fixed, we require our fits to find it as well. 

\section{The chaotic case}\label{secIV}
In this section we show that our proposed model, Eq. \eqref{distribution}, can be used to describe the distribution of ratios for both GOE and GUE limits \textit{with less error} than the original distributions announced in \cite{ratios}. It avoids finite-size effects with $N$, and the error decays as a power law when the number of realizations $M$ is increased. GOE and GUE reflect the most common symmetries found in realistic physical systems \cite{Modak2014,Moore2016}. 

The Wigner-like surmises that we take as theoretical expected results are %for each of the regularity and symmetry classes 
 the simple Poisson result and 
\begin{equation}\label{teoricas2}
P_{W}(r;\beta=1,2,4)\equiv\frac{1}{Z_{\beta}}\frac{(r+r^{2})^{\beta}}{(1+r+r^{2})^{1+3\beta/2}},
\end{equation}
where $\beta\in\{1,2,4\}$ correspond to GOE, GUE, and GSE, and $Z_{\beta}$ is as in \cite{ratios}. They were explicitly derived by exact calculation for $3\times3$ random matrices, and its applicability has been extended to arbitrary dimensions. Our model perfectly reproduces the $3\times 3$ statistics. However, the latter is not the most relevant scenario for many applications.

\subsection{The GOE limit}
\begin{figure}[h]
\hspace*{-0.5cm}
\includegraphics[width=0.38\textwidth]{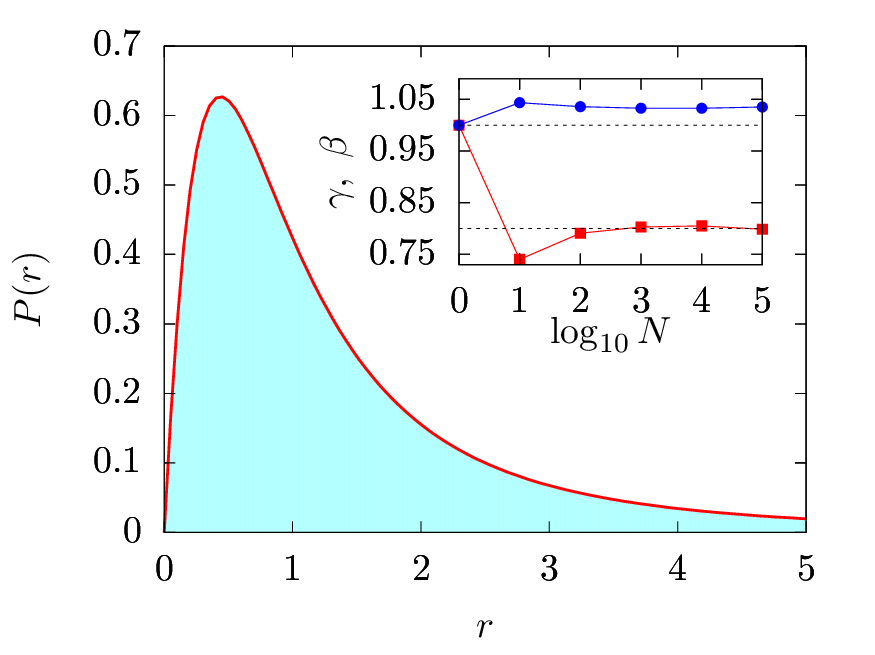}
\caption{(Color online) $P(r)$ calculated with an ensemble of $M=10^{5}$ {GOE} matrices giving rise to $N=10^{3}$ ratios each (blue histogram), and non-linear fit of our model $P_{\gamma\beta}(r)$, Eq. \eqref{distribution} (red, solid line). Bin size has been chosen $\delta r=0.005$. Inset: scaling of $\gamma$ and $\beta$ for GOE. Black dashed lines represent $\gamma=4/5$ and $\beta=1$. The same total number of ratios, $10^8$, has been used. }
\label{GOE106}
\end{figure}

To determine the accuracy of our model, we now examine an ensemble of  GOE random matrices of dimension $N\gg1$. The  $P(r)$ for this situation and fit of our model can be found in Fig. \ref{GOE106} for $N=10^{3}$ and $M=10^{5}$ realizations. These are visually indistinguishable. The non-linear fit of Eq. \eqref{distribution} produces $\beta=1.033(4)$ and $\gamma=0.8036(9)$. The limits proposed in Table \ref{info} has been chosen in accordance with this result. Both $\beta$ and $\gamma$ depend on the system size and, consequently, it is not consistent to set the exact results of our fit, but a simplified version. A scaling analysis of $\gamma$ and $\beta$ for GOE matrix sizes up to $N=10^{5}$ \cite{explanation4} is shown in the inset of Fig. \ref{GOE106}. For $N=3$, we obtain $\gamma=1$. It then departs from the $3\times 3$ result, and reaches an asymptotic value $\gamma\approx 4/5$ for $N=10^{3}$, at which it remains stable even for $N=10^{5}$. Fluctuations of $\beta$ in the entire range are very small overall, so we set $\beta=1$ for GOE regardless of $N$. 

\begin{figure}[htbp]
\centering
\hspace*{-0.8cm}
\includegraphics[width=0.40\textwidth]{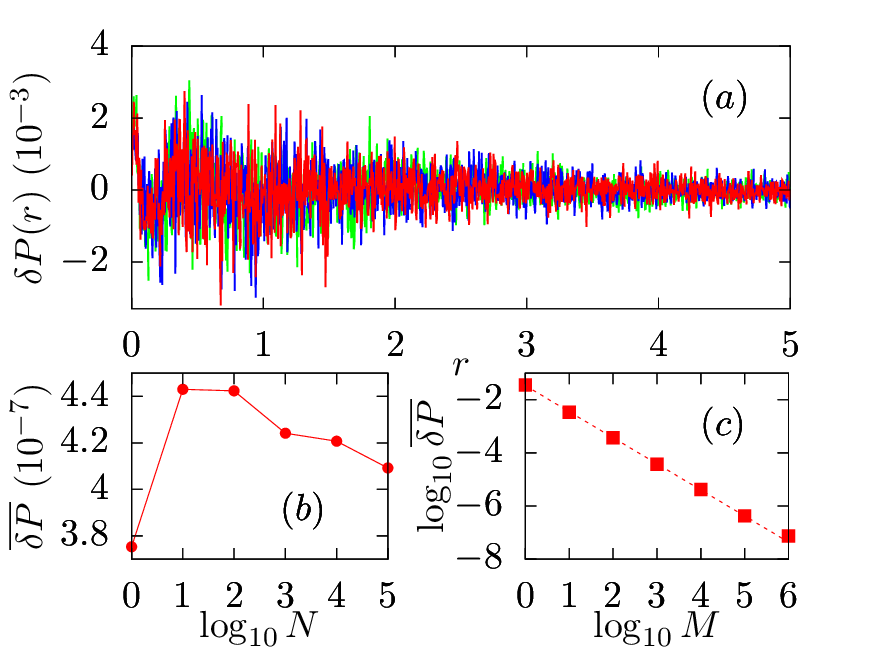}
\caption{(Color online) Panel $(a)$: Difference $\delta P_{\gamma\beta}(r)$, Eq. \eqref{differences}, between simulated histogram coming from $M\in\{10^{7},10^{5},10^{3}\}$ {GOE} random matrices providing a number of $N\in\{10,10^{3},10^{5}\}$ ratios each and our interpolating surmise $P_{\gamma\beta}(r)$, Eq. \eqref{distribution} (green, blue, and red lines). Panel $(b)$: scaling of $\overline{\delta P}$, Eq. \eqref{meandifferences}, with $N$. The same number of ratios, $10^{8}$, has been used. Panel $(c)$: Scaling of $\overline{\delta P}$ with $M$ for $N=10^{3}$. Dashed lines represent the best linear fit $\overline{\delta P}\propto M^{-0.982(5)}$.}
\label{errorsGOE106}
\end{figure}

We now assess the error that our estimate produces. %how good our estimate is in terms of difference between the simulated data and the results predicted by the theoretical value given by the Wigner-like surmise \cite{ratios}.
%We denote the Wigner-like surmise
%\begin{equation}\label{teoricas2}
%P_{W}(r;\beta=1,2,4)\equiv\frac{1}{Z_{\beta}}\frac{(r+r^{2})^{\beta}}{(1+r+r^{2})^{1+3\beta/2}},
%\end{equation}
%where $\beta\in\{1,2,4\}$ correspond to GOE, GUE, and GSE, respectively and $Z_{\beta}$ is as in \cite{ratios}.
Here we are interested in $P_{W}(r;\beta=1)$. We calculate 
\begin{equation}\label{differences}
    \delta P_{i}(r)\equiv  P_{{H}}(r)-P_{i}(r),
\end{equation}
where $P_{H}(r)$ denotes the distribution of ratios of given by the numerical histogram, and $P_{i}(r)$, with $i\in\{W,\gamma\beta\}$ represents the Wigner-like distribution, Eq. \eqref{teoricas2}, and our model, $P_{\gamma\beta}(r)$. The results for $\delta P_{\gamma\beta}(r)$ are plotted in panel $(a)$ of Fig. \ref{errorsGOE106} for $N\in\{10,10^{3},10^{5}\}$. The errors, very small in all cases, seem to behave like a random noise with no structure: $\delta P_{\gamma\beta}(r)$ only seems higher where $P(r)$ is too. In panel $(b)$ of the same figure, we display a scaling of the mean error, \begin{equation}\label{meandifferences}\overline{\delta P}\equiv \frac{1}{n}\sum_{j=1}^{n}|P_{H}(r_{j})-P_{\gamma\beta}(r_{j})|^{2},
\end{equation} where the integer $n\in\mathbb{N}$ is the total number of bins. The error can no longer be described by formulae such as those in \cite{ratios}, and remains approximately constant, with very small fluctuations, irrespective of $N$. Since Eq. \eqref{distribution} does not suffer from finite-size errors, in contrast to the $3\times 3$ surmise, it can be used in studies where the system size plays a relevant role.  On the scale of $\delta P_{W}(r;\beta=1)\sim 10^{-2}$ (not shown; see \cite{ratios} instead), the error produced by our model is quite negligible, of order $\delta P_{\gamma\beta}(r)\sim 10^{-3}$, $0\leq r\lesssim2$. In the domain $2\lesssim r \leq 5$, it becomes even smaller. Therefore, our model reproduces the histogram values with much more accuracy than the theoretical surmise. 

Our formula is expected to be exact at $\gamma(\beta=1)=1$ when the number of realizations $M\to\infty$ and the number of bins $n\to\infty$. Thus, the error at $N=3$ can only arise because these limits are not reached. In panel $(c)$ of Fig. \ref{errorsGOE106}, we display the scaling of $\overline{\delta P}$ with $M$ for a representative choice of the matrix size, $N=10^3$ \cite{explanation5}. We find almost perfect exponential decay of the form $\overline{\delta P}\propto M^{-0.982(5)}$, which is compatible with our previous statement ---that the error appears to be random and vanishing as the total number of ratios tends to infinity. 

\subsection{The GUE limit}
\begin{figure}[h]
\hspace*{-0.5cm}
\includegraphics[width=0.38\textwidth]{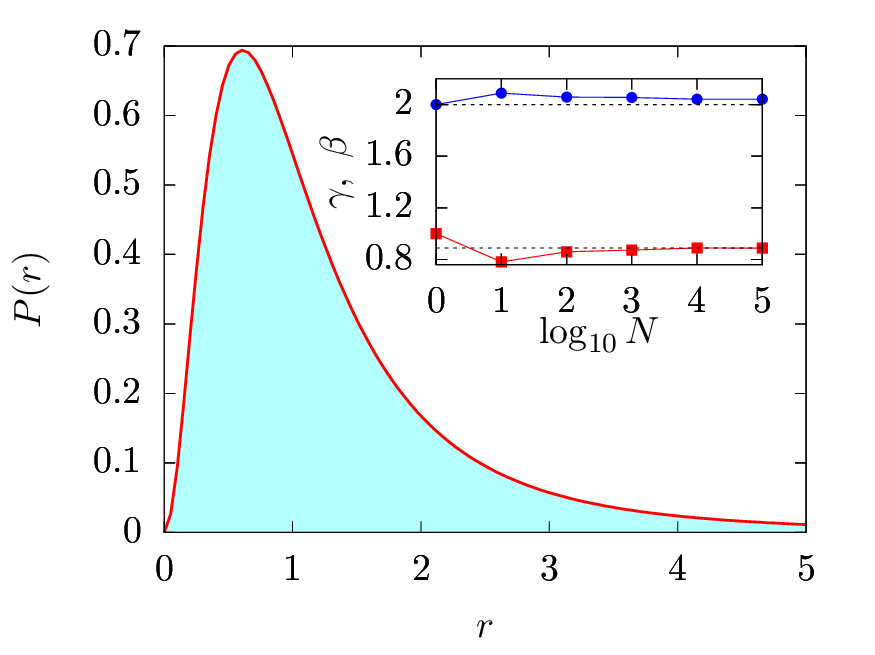}
\caption{(Color online) $P(r)$ calculated with an ensemble of $M=10^{5}$ {GUE} matrices giving rise to $N=10^{3}$ ratios each (blue histogram), and non-linear fit of our model, $P_{\gamma\beta}(r)$ Eq. \eqref{distribution} (red, solid line). Bin size has been taken $\delta r=0.005$. Inset: scaling of $\gamma$ and $\beta$ for GUE. Black dashed lines represent $\gamma=8/9$ and $\beta=2$. The same number of ratios, $10^8$, has been used.  }
\label{GUE106}
\end{figure}

\begin{figure}[h]
\hspace*{-0.5cm}
\includegraphics[width=0.40\textwidth]{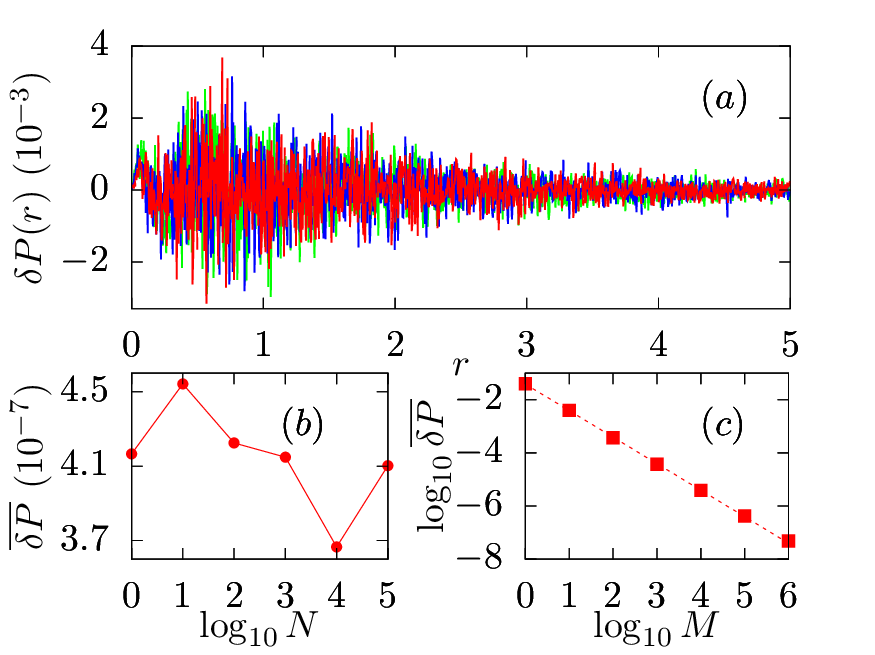}
\caption{(Color online)  Panel $(a)$: Difference $\delta P_{\gamma\beta}(r)$, Eq. \eqref{differences}, between simulated histogram coming from $M\in\{10^{7},10^{5},10^{3}\}$ {GUE} random matrices providing a number of $N\in\{10,10^{3},10^{5}\}$ ratios each and our interpolating surmise $P_{\gamma\beta}(r)$, Eq. \eqref{distribution} (green, blue, and red lines). Panel $(b)$: scaling of $\overline{\delta P}$, Eq. \eqref{meandifferences}, with $N$. The same number of ratios, $10^{8}$, has been used. Panel $(c)$: Scaling of $\overline{\delta P}$ with $M$ for $N=10^{3}$. Dashed lines represent the best linear fit $\overline{\delta P}\propto M^{-0.998(5)}$. }
\label{errorsGUE106}
\end{figure}

Quantum chaotic systems can also exhibit invariance under unitary transformations. We now test our interpolating model at the GUE limit and analyze the results it yields  compared with the theoretical value $P_{W}(r;\beta=2)$, Eq. \eqref{teoricas2}. Our findings are now displayed in Fig. \ref{GUE106} and Fig. \ref{errorsGUE106}.

In Fig. \ref{GUE106} we observe a perfect match of our model and the histogram of the simulated ratios. The double-parameter fit provides values $\beta=2.049(1)$ and $\gamma=0.879(1)$. Again, the ansatz proposed in Table \ref{info} is a simplified version of this last result. In the inset, we observe the scaling of $\gamma$ and $\beta$ with $N$: $\gamma$ departs from the Wigner surmise, $\gamma=1$, at $N=3$ and plateaus at $\gamma\approx 8/9$ for $N=10^{3}$. Changes in $\beta$ are again quite irrelevant. In panel $(a)$ of Fig. \ref{errorsGUE106} we encounter the same qualitative behavior of $\delta P_{\gamma\beta}(r)$ for GUE.  In panel $(b)$ of the same figure, we show $\overline{\delta P}$ for GUE. It is small and very similar for all values of $N$ considered, deviating very little from the result at $N=3$. Finally, in panel $(c)$ we show the scaling with the number of realizations, $M$, for $N=10^3$, which yields the best linear-fit result $\overline{\delta P}\propto M^{-0.998(5)}$, very similar to that of GOE. In summary, Eq. \eqref{distribution} has an error that remains approximately constant as $N$ is varied and reduces the discrepancy with numerical evidence with respect to the theoretical expectation, highly sensitive to $N$. Our results are compatible with the error being random, power-law decaying as $M\to\infty$, and very similar to the error at $N=3$. 

\section{Crossovers from integrability to chaos}\label{SecV}
We now focus on transitions between full chaos and the integrable regime, from RMT ensembles to real physical systems. Some systems, such as non-KAM or pseudointegrable systems, are not considered here.  %We also address the question of the universality of these transitions and find that the uniqueness of any interpolating formula of our kind is not, and in fact will never be, possible. 

Our \textit{main conclusion} is that the transition from chaos to regularity in different systems requires different parametrizations of $\gamma(\beta)$. Although this is only numerically demonstrated for a set of systems, it is enough to qualitatively observe that the crossovers will generally be strongly system-dependent. Contrary to what happens in classical mechanics, where the ratio of chaotic phase space can be used as a proper measure of chaos, \textit{this result illustrates the serious difficulty of defining a proper measure of quantum chaos by means of spectral statistics.} Later, in Sec. \ref{secVI}, we propose an ansatz to rely on the generalized Dyson index $\beta$ for this purpose.

%For quantum systems with a classical analogue, this is a consequence of the phase space features. This seriously hinders attempts to derive general interpolating formulas that hold with generality.

%Our analysis results in an ansatz proposal for the functional form of $\gamma=\gamma(\beta)$ that depends on the kind of transition to which it can be applied. This affords a convenient expression that can be used in quite general scenarios, so long as the amount of available data does not entail a huge large-$N$ limit. It must be emphasized that this represents the usual situation for most experimental investigations in complex systems, such as atoms or nuclei, as well as for many theoretical studies where the extraction of eigenlevels, of vital importance in the characterization of regularity classes, becomes seriously hindered by the exponential size scales of the Hilbert space.

\subsection{Description of models}
We first introduce the models that we use in this part. Eigenlevels have been obtained by full diagonalization in all cases.

$(i)$ \textit{Poisson to GOE transition in RMT}. This is the simplest transition one can consider. It does not correspond to any particular physical system. It consists in explicitly generating the Poisson and GOE limits, whose matrices in each realization we denote $\mathcal{H}_{P}$ and $\mathcal{H}_{G}$, respectively, and then building up the mixture, dependent on the continuous chaoticity parameter $\lambda$. This accomplished by the usual convex sum
\begin{equation}\label{modelp2goe}
    \mathcal{H}(\lambda)\equiv \lambda\mathcal{H}_{G}+(1-\lambda)\mathcal{H}_{P},\,\,\,\lambda\in[0,1],
\end{equation}
with limiting values $\mathcal{H}(\lambda=0)=\mathcal{H}_{P}$ and $\mathcal{H}(\lambda=1)=\mathcal{H}_{G}$. Other models for this transition are also possible, but this is perhaps the easiest one \cite{explanation2}. For our simulations, we have chosen the perturbation parameter $\lambda\in\{1.22^{q-1}\times 20^{-6}\}_{q=1}^{51}$. Here, the limiting value $\lambda=1$ is never reached because it is not essential for our purposes: the system in fact becomes chaotic for $\lambda\ll1$. A number of realizations $M=2000$ has been performed and the matrix size of each of them is $N=1716$.

$(ii)$ \textit{Poisson to GOE transition in the Gaussian $\mathbf{\beta}-$ensemble}. 
Also known as the Continuous Gaussian Ensemble, this generalization of the classical Gaussian ensembles was in its origins studied as a theoretical joint eigenvalue distribution with applications, for instance, in lattice gas theory \cite{lattgas2}. This eigenvalue distribution can be derived from an ensemble of random matrices \cite{dimitriu}. The Gaussian $\beta-$ensemble has since been used for various purposes \cite{powerspectrum,lecaer}. It has been proposed as a model to describe short-range statistics of the many-body to localized phase transition \cite{prl2019}. The ensemble essentially consists of tridiagonal, real, and symmetric matrices whose entries are classical random variables, these being normal, $\mathcal{N}(\mu,\sigma)$ with $\mu$ being its mean and $\sigma$ its standard deviation, and chi, $\chi_{k}\equiv \sqrt{\chi_{k}^{2}}$ with $k\in\mathbb{R}_{+}\cup\{0\}$ denoting a continuous, non-negative number of degrees of freedom. The matrix elements of the model $\mathcal{H}_{i,j}\equiv(\mathcal{H})_{i,j}$ are 
\begin{equation}
    \mathcal{H}_{i,i}\sim\mathcal{N}\left(0,\sqrt{\frac{1}{2\lambda}}\right), \,\,\,\forall i\in\{1,2,\dots,N\},
    \end{equation}
    and
    \begin{equation}
    \mathcal{H}_{i+1,i}=\mathcal{H}_{i,i+1}\sim\sqrt{\frac{1}{4\lambda}}\chi_{(N-i+1)\beta},\,\,\,\forall i\in\{1,2,\dots,N-1\},
\end{equation}
with $\lambda,\in\mathbb{R}_{+}$, $\beta\in[0,+\infty)$ being free parameters. The values $\beta=0,1,2,4$ correspond to Poisson, GOE, GUE, and GSE, respectively \cite{explanation1}. For consistency, here the convention that $\chi_{0}\equiv 0$ is assumed. For our simulations, we have made the simple choice $\lambda=1$ and $\beta\in\{0.02(q-1)\}_{q=1}^{51}$. We have averaged over $M=2000$ realizations, and the matrices size is $N=1716$.

$(iii)$ \textit{Poisson to GOE transition in a Heisenberg XXZ spin-${1/2}$ chain}.
Disordered interacting spin$-1/2$ chains have been used as models for quantum computers, magnetic compounds, and have been simulated in optical lattices \cite{Bloch2008,Simon2011,Trotzky2012}. Our model has been shown to transit from integrability to chaos, for instance, with the NNSD \cite{lfsantos,serbyn16,prl2019} or the $\delta_{n}$ \cite{us}. The Hamiltonian of the model is given by 
\begin{equation}\label{hamiltonian}
\mathcal{H}=\sum_{n=1}^{L}\omega_{n}\hat{S}_{n}^{z}+\sum_{n=1}^{L-1}J\hat{S}_{n}\cdot \hat{S}_{n+1},
\end{equation}
where $L$ is the number of sites and $\hat{S}_{n}\equiv \vec{\sigma}_{n}/2$ are the spin operators located at site $n$ with $\vec{\sigma}_{n}$ being the Pauli spin matrices at that site. The first term in Eq. \eqref{hamiltonian} describes effects of a static magnetic field in the $z$-direction. Each $\omega_{n}$ is a random variable distributed uniformly over $[-\omega,\omega]$. Two possible couplings between the nearest neighbor spins are described by the last term of Eq. \eqref{hamiltonian}. The first one is simply the diagonal Ising interaction, while the second is the off-diagonal flip-flop term, which is responsible for excitation propagation in the chain. The chain is taken to be isotropic.% since $J$ is a constant that quantifies both the coupling strength between the Ising interaction and that of flip-flop term. 

For our simulation, we
have taken $J=1$, $\hbar\equiv1$, and $L=13$. Periodic boundary conditions are used to minimize finite-size effects. We have generated $50$ different cases, with $\omega\in\{0.2q\}_{q=1}^{50}$. The transition is believed to be completed around $\omega\sim 3.6$ \cite{Luitz15}. Since $[H,\hat{S}_{z}]=0$, we consider
the sector $\hat{S}_{z}=-1/2$. The dimension of the
Hilbert space is $N=\binom{13}{6}=1716$, and we have simulated $M=2000$
realizations.

$(iv)$ \textit{Poisson to GOE transition in a Gaudin elliptic model}. This model is based on long-range interactions between spin-1/2 magnets. It is the most general of a family of exactly solvable models derived from a generalized Gaudin algebra \cite{gaudin1976}, which includes the Bardeen-Cooper-Schrieffe (BCS), the Suhl-Matthias-Walker, the Lipkin-Meshkov-Glick, the generalized Dicke, and nuclear interacting boson models, to quote but a few \cite{ortiz2005}. This familiy includes the XXX (rational), the XXZ (trigonometric-hyperbolic), and the XYZ (elliptic) classes. The rational one is known to coincide with the classical BCS mean-field solution in the thermodynamic limit \cite{dukelsky2004}. Here, we work with its XYZ version, which has been previously studied in \cite{gaudinarmando}, and can be written 
\begin{equation}\label{gaudin1}
    \mathcal{H}=\sum_{i=1}^{d}\epsilon_{i}R_{i},
\end{equation}
where $d$ is the number of spins, $\epsilon_{i}$ are free parameters, and $R_{i}$ are two-spin operators of the form 
\begin{equation}\label{gaudin2}
R_{i}\equiv \sum_{i<j}^{d}\widetilde{X}_{i,j}\sigma_{i}^{x}\sigma_{j}^{x}+\widetilde{Y}_{i,j}\sigma_{i}^{y}\sigma_{j}^{y}+\widetilde{Z}_{i,j}\sigma_{i}^{z}\sigma_{j}^{z},
\end{equation}
with $\sigma^{x},\sigma^{y},\sigma^{z}$ being the Pauli matrices. The matrices $\widetilde{X}$, $\widetilde{Y}$, and $\widetilde{Z}$ can be chosen to induce a complete transition from integrability to fully developed chaos via a single-parametric perturbation, $\alpha$. Following the proposal in \cite{gaudinarmando} we choose 
\begin{equation}\begin{split}\label{gaudin3}
    & \widetilde{X}_{j,k}=(\cos\alpha)X_{j,k}+(\sin\alpha)A_{j,k},\\ & \widetilde{Y}_{j,k}=(\cos\alpha)Y_{j,k}+(\sin\alpha)B_{j,k},\\ 
    & \widetilde{Z}_{j,k}=(\cos\alpha)Z_{j,k}+(\sin\alpha)C_{j,k}.
\end{split}\end{equation}
The $X,Y,Z$ matrices are written
\begin{equation}\begin{split}\label{gaudin4}
    & X_{j,k}=\frac{1+\kappa\, \textrm{sn}^{2}(z_{j}-z_{k})}{\textrm{sn}(z_{j}-z_{k})},\\
    & Y_{j,k}=\frac{1-\kappa\, \textrm{sn}^{2}(z_{j}-z_{k})}{\textrm{sn}(z_{j}-z_{k})},\\
   &  Z_{j,k}=\frac{\textrm{cn}(z_{j}-z_{k})\textrm{dn}(z_{j}-z_{k})}{\textrm{sn}(z_{j}-z_{k})},
\end{split}\end{equation}
where $z_{j}\in\mathbb{R}$, $j\in\{1,\dots,N\}$, are free parameters, $\textrm{sn}(x)\equiv\textrm{sn}(x,\kappa)$ is the Jacobi elliptic function of modulus $\kappa\in[0,1]$, and $\textrm{cn}$ and $\textrm{dn}$ are related by $\textrm{d}\,\textrm{sn}(x)/\textrm{d}x=\textrm{cn}(x)\textrm{dn}(x)$. They give rise to an integrable model, which can be solved by the Bethe ansatz \cite{gould02}. If $\alpha=0$, all the $R$ matrices commute pairwise, $[R_i, R_j]=0$, $\forall i\neq j$, and thus the system has as many integrals of motion as degrees of freedom. 

The remaining set of matrices are used to break the integrability of the model. They are chosen as
\begin{equation}\begin{split}\label{gaudin5}
    A_{j,k}=\mu+\sigma\cos\left[\sqrt{2\lambda}(\omega_{j}-\omega_{k})\right],\\
    B_{j,k}=\mu+\sigma\cos\left[\sqrt{3\lambda}(\omega_{j}-\omega_{k})\right],\\
    C_{j,k}=\mu+\sigma\cos\left[\sqrt{5\lambda}(\omega_{j}-\omega_{k})\right],
\end{split}\end{equation}
with $\lambda,\omega_{i}\in\mathbb{R}$ being free parameters; $\mu$  the average of all matrix elements of $A$, $B$, and $C$, and $\sigma$  the standard deviation. The transiting parameter $\alpha\in[0,\pi/2]$ is such that for $\alpha=0$ the system is completely regular and for $\alpha=\pi/2$ it is completely chaotic. We simulate $M=3000$ realizations of chains with $d=11$ spins, each one giving rise to a system of dimension $N=2^{10}=1024$, due to the presence of a discrete symmetry. As it has been shown \cite{gaudinarmando} that the transition to chaos is completed around $\alpha \sim \pi/4$, we choose $\alpha\in\{\pi q /200\}_{q=1}^{60}$.

$(v)$ \textit{Poisson to GUE transition}. We generate the crossover Hamiltonian as 
\begin{equation}\label{hamiltonianp2gue}
    \mathcal{H}(\lambda)\equiv \lambda\mathcal{H}_{\textrm{GUE}}+(1-\lambda)\mathcal{H}_{\textrm{P}},
\end{equation} so that $\mathcal{H}(\lambda=0)=\mathcal{H}_{\textrm{P}}$ and $\mathcal{H}(\lambda=1)=\mathcal{H}_{\textrm{GUE}}$. Here, we have simulated $M=2000$ realizations consisting of matrices of order $N=1716$ for each value of $\lambda\in\{1.34^{q}\times10^{-6}\}_{q=1}^{50}$. The crossover happens very fast as $\lambda$ is increased. %After obtaining the histogram $P(r)$ we perform a non-linear fit for our ansatz Eq. \eqref{ansatzp2gue}, and produce the transition that is now shown as visual proof in Fig. \ref{p2gue}.  Best fit results are also given in Table \ref{tablep2gue}. The limiting values of $\beta$ are in accordance with our expectation.

$(vi)$ \textit{GOE to GUE transition}. The Hamiltonian is now instead \begin{equation}\label{hamiltoniangoe2gue}
    \mathcal{H}(\lambda)\equiv \lambda\mathcal{H}_{\textrm{GUE}}+(1-\lambda)\mathcal{H}_{\textrm{GOE}},
\end{equation} so that $\mathcal{H}(\lambda=0)=\mathcal{H}_{\textrm{GOE}}$ and $\mathcal{H}(\lambda=1)=\mathcal{H}_{\textrm{GUE}}$. We have simulated  $\lambda\in\{1.34^{q}\times10^{-6}\}_{q=1}^{50}$, $M=2000$ and $N=1716$.

\subsection{Universality of crossovers} 
\begin{figure}[h]
\hspace*{-0.3cm}
\begin{tabular}{c}
\includegraphics[width=0.39\textwidth]{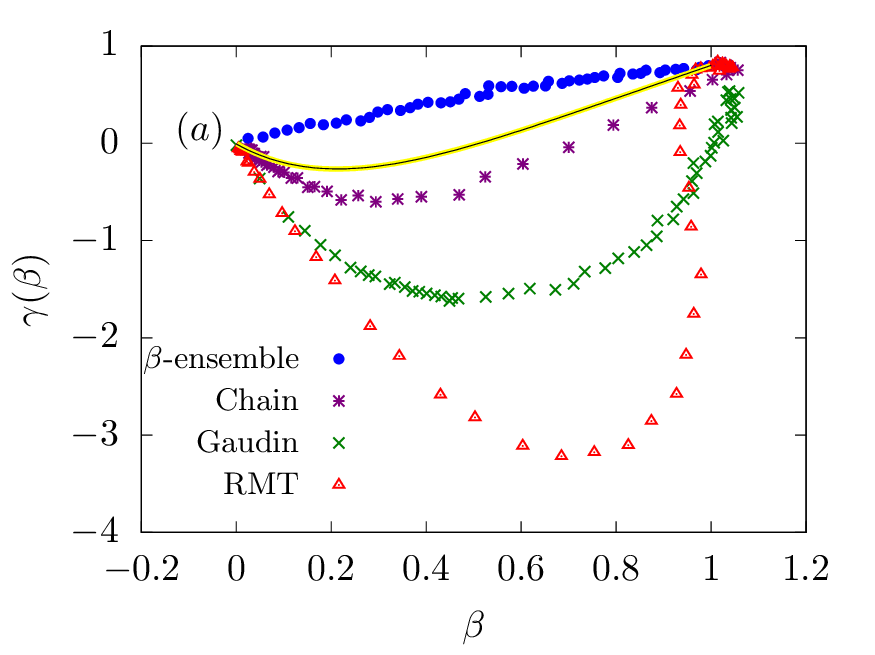}\\ 
\includegraphics[width=0.39\textwidth]{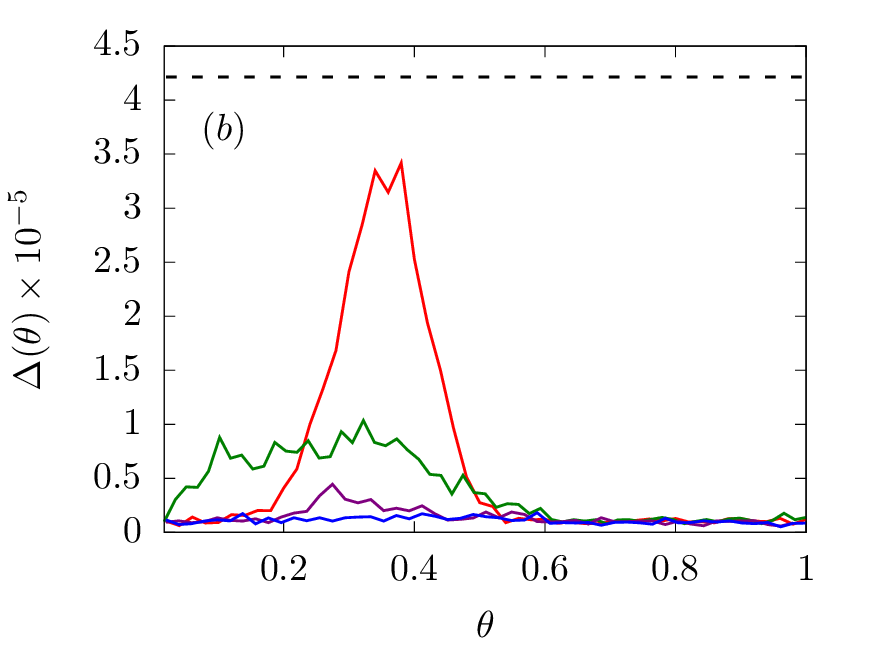}
\end{tabular}
\caption{(Color online) Panel $(a)$: Best non-linear double-fit result $\gamma=\gamma(\beta)$ for the distribution, simulated for the Poisson-GOE crossovers. Plotted against them are the $\gamma(\beta)$ obtained from information entropy (yellow line) and our proposed ansatz for the Poisson-GOE crossover (black line). Panel $(b)$: Eq. \eqref{delta} for the corresponding curves. Last (GOE) value of the $\beta-$ensemble (black, dashed line) sets an upper limit. Bin sizes are $\delta r=0.05$. }
\label{gammabeta}
\end{figure}

One of the questions that needs to be addressed about our transiting model Eq. \eqref{distribution} is the existence of a functional form for $\gamma=\gamma(\beta)$ that might be applicable to any generic physical system.  The simulated data provides all we need to construct the distribution of the ratio of two consecutive level spacings, Eq. \eqref{ratio}, via the diagonalization of the Hamiltonian of each transition. For each one of them, and for each value of the perturbative parameter therein, we proceed with a non-linear fit of our equation $P(r;\beta,\gamma(\beta))$ to the distribution given by the histograms, $P_{H}(r)$. We take $\delta r=0.05$ for all cases, so results can be put in comparison.  %This process involves a three-parameter fit, since for each case we will be needing $\beta$, $\gamma$, and, for completeness, we will also require the normalization constant $C_{\beta}$ to be found, even though it may still be exactly calculated once the other two parameters are known via the normalization condition. 
These are shown in Fig. \ref{gammabeta}, which is the second \textit{main result} of our work.

Panel $(a)$ of Fig. \ref{gammabeta} refutes the possibility of a universal transiting formula for the ratios. Indeed, one such expression would need to describe the behavior of systems that exhibit not only quantitatively but also qualitatively very different crossovers. As can be seen, there is no easy way to characterize these four curves at the same time. We observe two qualitatively different curves: the transition associated to the system $(ii)$, that is, the $\beta-$ensemble, is slightly concave; contrarily, the curves that correspond to the systems $(i)$, $(iii)$, and $(iv)$, although not so much quantitatively, share a qualitative commonality in that all of them are convex. These two groups of functions are obviously mutually exclusive. In addition, once the general shape has been accounted for, these last three curves have nothing in common. The differences in the form of $\gamma(\beta)$ between the $\beta-$ensemble and the XXZ spin$-1/2$ chain model are incidentally in concert with the results of \cite{piotr2019}. In passing we note that although $\gamma(\beta)$ is multivalued for $0.9\lesssim \beta\lesssim1$, this is entirely due to fluctuations and should not be taken seriously. %In addition, we have plotted an average of all four curves, which we show with a black dashed line. Finally, we plot with a grey solid line the ansatz for the Poisson-GOE crossover given in Table \ref{ansatzs}. 
 The curves $\gamma(\beta)$ are visibly quite difficult to parametrize in terms of simple functions, let alone a family of functions.

In order to determine the goodness of fit of Eq. \eqref{distribution}, we calculate the difference between the best fit and the numerical histograms. %To this end, we consider a partition of the interval under consideration $[0,5]$ given by \begin{equation}\label{partition}\mathcal{P}\left([0,5]\right)\equiv \left\{[\delta r(i-1),\delta r(i))\right\}_{i=1}^{n},\end{equation} where $\delta r$ is, as usual, the bin size in the histogram and is such that $\delta r\times n=5.$ Then the number $n\in\mathbb{N}$ can be interpreted as the \textit{number of bins} employed in our analysis. For each interval in the partition Eq. \eqref{partition}, we consider the $i$-th mid-point $r_{i}\equiv (2i-1)\delta r/2\in[0,5]$, $i\in\{1,2,\dots,n\}$. 
If we let $P^{(q)}(r)$ denote the distribution of the ratios for the $q$-th value of the transition parameter, then 
\begin{equation}\label{delta}
\Delta(q)\equiv \frac{1}{n}\sum_{j=1}^{n}\left|P_{H}^{(q)}(r_{j})-P_{\gamma\beta}^{(q)}(r_{j})\right|^{2},\,\,\forall q\in\{1,2,\dots,q_{\max}\},
\end{equation}
 represents the squared difference between the histogram values and the distribution fits at each point $r_{i}$ averaged over the total number of bins. Here, we include all numbers of realizations. %This representation can be used to measure the discrepancy between simulated data in the transition and our models.
 Note that Eq. \eqref{delta} supplies results that are effectively independent of the number of bins. %When several numbers of realizations $M$ are taken into account to construct the distribution for each step of the transition, \textit{all} of them must be included in Eq. \eqref{delta}. 
 Since the number of parameters for the crossovers have been chosen slightly different depending on the system, we plot the results as a function of the \textit{normalized parameter} $
    \theta\equiv{q}/{q_{\max}}\in[0,1],\,\,\,q\in\{1,2,\dots,q_{\max}\},$
where $q_{\max}$ is the highest value of $q$ for each system.

In panel $(b)$ of Fig. \ref{gammabeta}, we show the results of Eq. \eqref{delta} applied to the fits displayed in panel $(a)$ of the same figure. %For reference, we also show, with a black, dashed line, the error between the numerical GOE and the theoretical expression $P_{W}(r;\beta=1)$, Eq. \eqref{teoricas2}.
The black, dashed line show that the double fitting to Eq. \eqref{distribution}, throughout the whole transition from integrability to chaos, produces less error than Eq. \eqref{teoricas2} in the GOE limit. 
%At the same time, it is also legitimate to consider the mechanism that induces such a transition: for systems I and III

It becomes apparent that there does not exist a unique $\gamma(\beta)$ that serves the ambitious purpose of entirely taking into account all possible systems with intermediate dynamics %---many such formulae could in principle be found, but \textit{they would be valid only for that system for which it was explicitly derived}. 
The one-variable choice for the structure of $\gamma(\beta)$ is incidentally reinforced by the results of Fig. \ref{gammabeta}, where a smooth plot is found for all four transitions. %It seems very doubtful that this result could be compatible with $\gamma$ depending on some other paremeter besides $\beta$. 

\begin{figure}[h]
\hspace*{-0.5cm}
\includegraphics[width=0.45\textwidth]{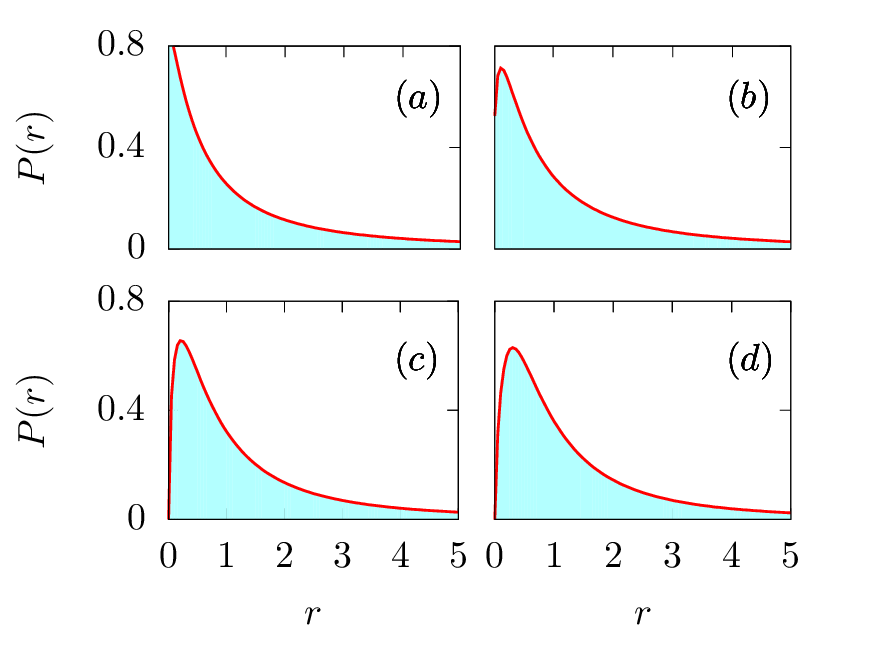}
\caption{(Color online) $P(r)$ for the Heisenberg spin-$1/2$ XXZ chain model with $N=1716$, $M=2000$, and $\delta r=0.05$ (blue, filled histogram), with the best non-linear fits of $P_{\gamma\beta}(r)$, Eq. \eqref{distribution} (red, solid line). The values of the transition parameter for panels $(a)-(d)$ are $\omega\in\{5.0,3.0,2.4,2.0\}$, respectively. }
\label{gaudin}
\end{figure}

In Fig. \ref{gaudin}, we show how our general surmise Eq. \eqref{distribution} works very well to describe crossovers with high enough statistics. This is exemplified by means of the Heisenberg spin$-1/2$ chain. Quantitative numerical results for the best non-linear double-fit are gathered in Table \ref{ansatzsgaudin}, where we observe that $\beta$ behaves as a monotonically increasing smooth function exhibiting very reasonable errors.

\begin{figure*}[t]
\hspace*{-0.5cm}
\includegraphics[width=0.85\textwidth]{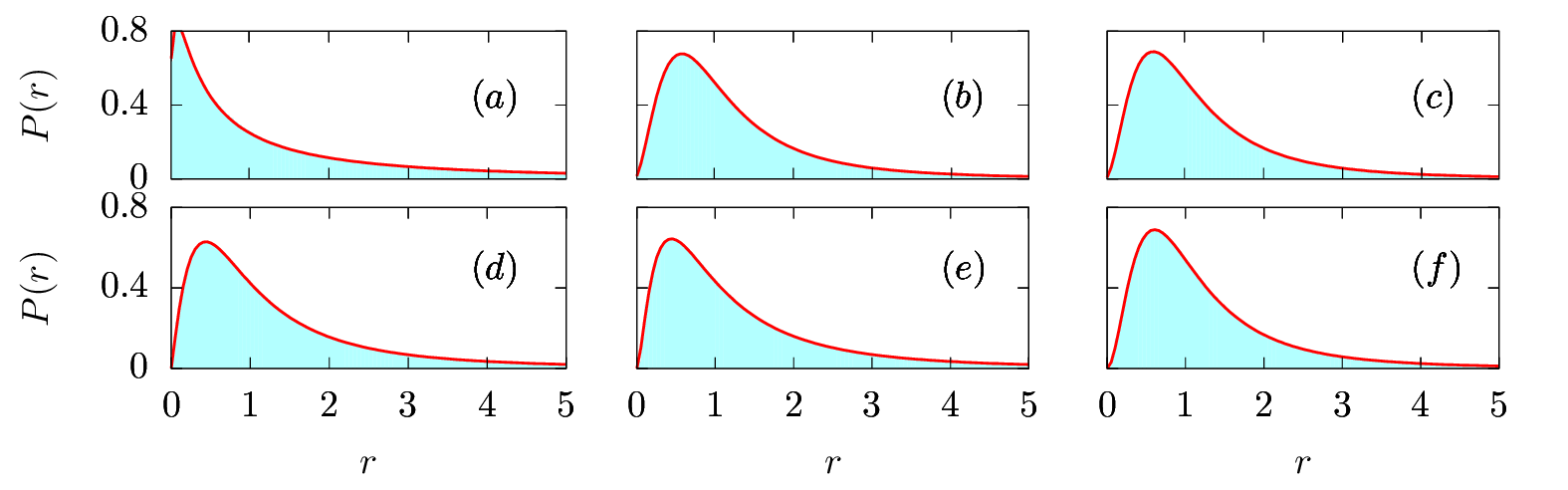}
\caption{(Color online) $P(r)$ (blue histograms) and best non-linear fits of $P_{\gamma\beta}(r)$, Eq. \eqref{distribution} (red lines). Panels $(a)-(c)$: the Poisson-GUE transition, with $\beta\in\{0.56,1.64,1.93\}$ and $\gamma\in\{-3.15,1.15,0.96\}$; panels $(d)-(f)$: the GOE-GUE transition, with $\beta\in\{1.12,1.66,1.94\}$ and $\gamma\in\{0.64,-0.12,-0.39\}$. In all cases, we simulate $M=2000$ realizations of matrices of dimension $N=1716$, and $\delta r=0.05$. }
\label{pgoegue}
\end{figure*}

The Poisson-GUE and GOE-GUE, $(v)$ and $(vi)$, crossovers are exemplified in Fig. \ref{pgoegue}. The values of $\beta$ of panels $(b)$ and $(e)$, and those of panels $(c)$ and $(f)$ are very similar. However, careful examination of the distributions shown reveals them to be quite different. Thus, our proposal can be used to differentiate several crossovers at the same value of $\beta$, contrarily to other past results, like a unique model for intermediate systems, e.g., the $\beta-$ensemble, or the Izrailev formula for the NNSD \cite{izrailev89}.

\begin{table}[h!]
\begin{center}
\setlength\extrarowheight{3pt}
 \begin{tabular}{||c c c||} 
 \hline
 Panel \hspace{0.5cm} & $\beta$ \hspace{0.5cm} & $\gamma$\\ [0.5ex] 
 \hline\hline
$(a)$ \hspace{0.5cm} & $0.1010(22)$ \hspace{0.5cm} & $-0.300(19)$  \hspace{0.5cm} \\[1ex]
 \hline
 $(b)$ \hspace{0.5cm} & $0.3898(44)$ \hspace{0.5cm} & $-0.549(25)$  \hspace{0.5cm} \\[1ex]
 \hline
 $(c)$ \hspace{0.5cm} & $0.6038(45)$ \hspace{0.5cm} &  $-0.211(18)$  \hspace{0.5cm}  \\ [1ex] 
 \hline
 $(d)$ \hspace{0.5cm} &  $0.7943(50)$ \hspace{0.5cm} &  $0.189(15)$ \hspace{0.5cm} \\ [1ex]
 \hline
\end{tabular}
\end{center}
\caption{Values and uncertainties of best non-linear double-fit of $P_{\gamma\beta}(r)$, Eq. \eqref{distribution}, referred to panels $(a)-(d)$ in Fig. \ref{gaudin}.}
\label{ansatzsgaudin}
\end{table}

\section{Ansatz proposal}\label{secVI}
A question that deserves exploration is the possibility of eliminating the dependence on $\gamma(\beta)$ in Eq. \eqref{distribution}. Two alternatives are initially possible: 

\textit{(i)} a double-fit $\gamma-\beta$ is always applicable, and 

\textit{(ii)} a compromise ansatz that requires no double-fitting is desirable. 

We will first analyze the difficulties that $(i)$ involves, and then conclude that $(ii)$ is the best option in terms of applicability, proposing such an ansatz. 

\begin{figure}[h]
\hspace*{-0.5cm}
\includegraphics[width=0.38\textwidth]{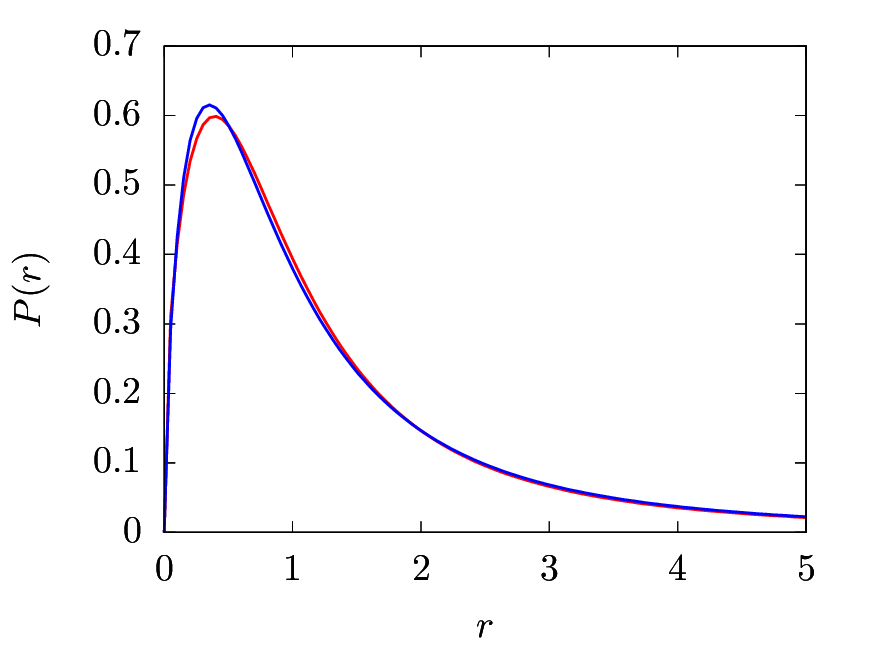}
\caption{(Color online) Distribution $P_{\gamma\beta}(r)$, Eq. \eqref{distribution}, for the values $\gamma(\beta=0.453)=1.643$ (red, solid line) and $\gamma(\beta=0.664)=0.887$ (blue, solid line). }
\label{illfit}
\end{figure}

$(i)$ \textit{Double-fitting shortcomings.} In Fig. \ref{illfit} we plot Eq. \eqref{distribution} with the choices $\gamma(\beta=0.453)=1.643$ and $\gamma(\beta=0.664)=0.887$. These values of $\beta\in(0,1)$ are associated with partially Poissonian or GOE dynamics. For the values of $\gamma$, we have made two very distinct choices. In conjunction with the values of $\beta$, it should reflect two very different dynamics. However, we find the curves to be almost indistinguishable, especially as $r\to0$ and $r\to\infty$. For the fit to differentiate between these two curves, we would need high statistics. This implies that casting Eq. \eqref{distribution} into a form for which a single-parameter fit suffices could be desirable. %since this would mean a much more robust result when one cannot obtain large statistics to analyze. %Note that in this case a double-parameter fit would produce ill-defined results and under no circumstances could these be taken as representative. 

\begin{figure}[h]
\hspace*{-0.5cm}
\includegraphics[width=0.38\textwidth]{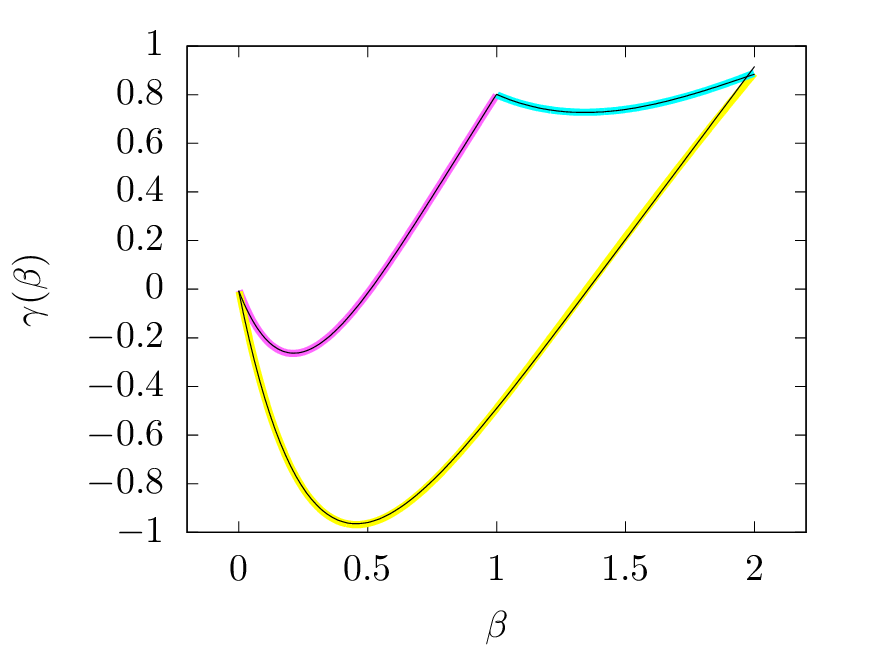}
\caption{(Color online) Values of $\gamma(\beta)$ that make the entropy change linear for the Poisson-GOE (magenta line), Poisson-GUE (yellow line), and GOE-GUE (cyan line) crossovers. Ansatzs, parametrized from $\gamma(\beta)$ and given in Tab. \ref{ansatzs}, are plotted with black lines. }
\label{entropyfig}
\end{figure}

$(ii)$ \textit{Choice for ansatzs.} In essence, we now seek to rewrite Eq. \eqref{distribution} so as to free it from the unknown $\gamma(\beta)$, that is, $P(r;\beta,\gamma(\beta))\mapsto P(r;\beta)$.
This transformation requires assigning $\gamma(\beta)$ an explicit form depending on $\beta$ alone. A physically relevant choice can be obtained taking into account the role that level repulsion plays in the uncertainty associated to both the NNSD and the $P(r)$. Due to level repulsion, chaotic spectra are more \textit{rigid} than integrable ones; that is, given the value of a particular energy level, the value of the next one is less uncertain in chaotic systems than it is in integrable ones. A proper measure of such an uncertainty can be obtained from the \textit{information entropy}, defined as \begin{equation}\label{entropy}S(\gamma,\beta)\equiv -\int_{0}^{\infty}\textrm{d}r\, P_{\gamma\beta}(r)\log P_{\gamma\beta}(r).\end{equation} 
Numerical values of $S$ at the limit ensembles are $S(\beta=0,\gamma=0)=2$, $S(\beta=1,\gamma=4/5)=1.45093$, and $S(\beta=2,\gamma=8/9)=1.17477$, respectively, confirming the previous statement. Hence, we propose for the compromise ansatz {\em the curve $\gamma(\beta)$ which linearly interpolates the information entropy between the limiting ensembles, as a function of $\beta$}. Then, the Dyson index $\beta$ can be understood as a measure of chaos: the larger the value of $\beta$, the less uncertain the corresponding $P(r)$ distribution. We find the linear interpolations $S_{\textrm{Poisson-GOE}}(\beta)=2-0.5407\beta$, $S_{\textrm{Poisson-GUE}}(\beta)=2-0.4126\beta$, and $S_{\textrm{GOE-GUE}}(\beta)=1.7271-0.2762\beta$. Numerically solving $S(\gamma,\beta)$ for $\gamma$ so that these interpolations hold affords the results in Fig. \ref{entropyfig}. The curves $\gamma(\beta)$ so obtained interestingly mimic those from particular physical systems in Fig. \ref{gammabeta}. We have then parametrized $\gamma(\beta)$ in terms of polynomials for each transition. This yields the ansatzs given in Table \ref{ansatzs}. 

As the transition from integrability to chaos is not universal, other possible choices for one-parametric transiting distributions are possible. It would be interesting to investigate in the future whether there exists a function $\gamma(\beta)$ that matches the Brody distribution for the NNSD. In Refs. \cite{Bhosale2018,Tekur2018b}, a scaling relation between the distribution of non-overlapping high-order ratios and that of the usual ratios of this work is presented. It is first postulated in \cite{Bhosale2018}, and the distribution for the Wishart ensemble at $\beta=1,2$ is there shown. In \cite{Tekur2018b}, the analysis is extended and applied to complex systems. It would be also interesting to investigate whether there exists a curve $\gamma(\beta)$ fulfilling such a scaling relation.

%It measures the uncertainity of the eigenlevel statistics. Numerically, for every $\beta$ we have chosen $\gamma$ so that $S$ evolves linearly from Poisson to GOE and GUE, and from GOE to GUE. . Because the possible outcomes for the level spacings decrease from Poisson to GUE, $S(\beta)$ is monotone decreasing. Then, $S(\beta)$ interpolates linearly between the ensembles as $S(0\leq\beta\leq1)=2-0.5407\beta$, $S(0\leq\beta\leq2)=2-0.412615\beta$, and $S(1\leq\beta\leq2)=1.72709-0.27616\beta$. This choice implies considering that as information uncertainty decreases, the degree of chaos increases linearly. 

\section{Conclusions}\label{SecVII}
The distribution of the ratio of consecutive level spacings, a short-range spectral statistic, has been gradually growing in popularity in the recent years since it does not require spectral unfolding, contrarily to the traditional NNSD and others. While theoretical expressions for the distribution $P(r)$ are known for the integrable case together with the three classical random ensembles, these were derived with $3\times 3$ random matrices. An analytical interpolating formula between these regularity classes and degrees of chaoticity remains unknown, despite the several attempts having been made in the past. 

In this paper, we have proposed an interpolating formula that depends on a single parameter, a generalized Dyson index, $\beta$, and on a function, $\gamma(\beta)$, that can be fitted as a second parameter. Our surmise fulfills the mathematical conditions that are imposed upon such a probability density function. This one-parameter expression is then found to reduce the discrepancy with exact both GOE and GUE simulations with respect to the theoretical Wigner-like results given in \cite{ratios}. Stringent scaling analysis allows us to conclude that our surmise does not suffer from finite size effects, producing a very similar error for all matrix sizes, and displaying an explicit power-law decay of the error as the number of realizations is increased. Our model also affords interesting corrections for $\langle r\rangle$ and $\langle \widetilde{r}\rangle$. 

We have analyzed whether the two-parameter dependence of our crossover model can be reduced to a single-parameter one in such a way that this still affords a good description of general crossovers. The answer has been found to be negative. By studying both RMT generated and real physical systems, we conclude that there cannot exist a universal expression for $\gamma(\beta)$ valid with absolute generality. This is due to the very particular features of the crossover for different systems. Because a two-parameter fit needs high statistics to be reliable, we use the information entropy to propose ansatzs of $\gamma(\beta)$ for the main crossovers, Poisson-GOE, Poisson-GUE, and GOE-GUE. %These expressions must differ from one another since, as we have already pointed out, there cannot be a general representation for this expression that guarantees universal success in describing crossovers.
%We show that, so long as the statistics is not huge, which is often the case in the vast majority of situations, employing our ansatzs does not result in significant worsening of the description afforded by the statistic $P(r)$ due to fluctuations, which justifies its applicability. 

%Finally, we show how our proposal can be used for less common transitions with success, such as the Poisson-GUE and the GOE-GUE crossovers. 
Interestingly, our results can be successfully used to distinguish different crossovers at the same value of $\beta$, meaning the particular kind of transition cannot be ignored. This reflects the versatility of our suggestions. 

In summary, we provide a generic formula for the ratio of consecutive level spacings that can be used to assess the degree of chaos for different symmetries (or mixture of them) under very general circumstances. 
\begin{acknowledgements}
This work has been supported by the Spanish Grants
Nos. FIS2015-63770-P (MINECO/ FEDER) and PGC2018-094180-B-I00 (MCIU/AEI/FEDER, EU).
\end{acknowledgements}


\begin{thebibliography}{100}
\bibitem{explanation4} Gaussian $\beta$-ensembles, known to afford exact representations for the standard limits, are used to reach such large sizes \cite{lattgas2,dimitriu}.
\bibitem{explanation5}
A compromise had to be reached about the value of $N$. Simulating $M\geq10^{5}$ realizations of matrices of dimension $N\geq10^{3}$ takes a huge computational time, so we decided to choose $N=10^{3}$.
\bibitem{explanation2}
In any event, we have meticulously checked that the model that implements the transition plays no significant role to our purposes and no qualitative modifications occur.
\end{thebibliography}
\end{document}